# Electron Beam Spectroscopy for Nanophotonics

Albert Polman[1,*], Mathieu Kociak[2] and F. Javier García de Abajo[3,4]

[1]Center for Nanophotonics, AMOLF, Amsterdam, the Netherlands

[2]Laboratoire de Physique des Solides, Université de Paris-Sud, Orsay, France

[3]ICFO-Institut de Ciencies Fotoniques, The Barcelona Institute of Science and Technology, Castelldefels (Barcelona), Spain

[4]ICREA-Institució Catalana de Reserca I Estudis Avançats, Barcelona, Spain

**Abstract**

Progress in electron-beam spectroscopies has recently enabled the study of optical excitations with combined space, energy and time resolution in the nanometer, millielectronvolt and femtosecond domain, thus providing unique access into nanophotonic structures and their detailed optical responses. These techniques rely on ~1-300 keV electron beams focused at the sample down to sub-nanometer spots, temporally compressed in wavepackets a few femtoseconds long, and in some cases controlled by ultrafast light pulses. The electrons undergo energy losses and gains, also giving rise to cathodoluminescence light emission, which are recorded to reveal the optical landscape along the beam path. This review portraits these advances, with a focus on coherent excitations, emphasizing the increasing level of control over the electron wave functions and ensuing applications in the study and technological use of optically resonant modes and polaritons in nanoparticles, 2D materials and engineered nanostructures.

# 1. Introduction: fundamentals of EELS and CL

Cathodoluminescence (CL) and electron energy-loss spectroscopy (EELS) have advanced in recent decades to arguably provide the best combination of space, energy and time resolutions for the structural and optical characterization of materials. In these techniques, an energetic electron beam is raster-scanned over the specimen using either a transmission electron microscope (TEM, featuring 30-300 keV electron beams and equipped with an electron analyzer for acquisition of EELS spectra, and optionally with an optical spectrometer for CL) or a scanning electron microscope (SEM, 1-30 keV for CL). The acquired CL/EELS infrared-to-UV spectral data are then correlated with morphological, and structural information derived from secondary electron (SE) images (mostly in SEMs) or the high-angle annular dark field (HAADF) signal (in TEMs), all taken on the same sample during the same measurement. Optionally, energy-dispersive X-ray spectroscopy (EDX), electron backscatter diffraction (EBSD) and high-energy core-level spectroscopy (in EELS) can simultaneously provide compositional and atomic-structure correlations.[1] Beyond traditional materials science applications, the last decade has witnessed the emergence of both EELS and CL as unique tools in research on the behavior of light at the nanoscale. Due to the unsurpassed spatial resolution offered by electron microscopes down to the atomic (TEM) or nanometer (SEM) scale, pixel-by-pixel comparisons can be made between CL/EELS images and compositional as well as morphological information at length scales that are small compared with the light wavelength. These nanoscale correlations are at the heart of the success of spatially-resolved EELS and CL in nanophotonics research.

Figure 1(a-d) shows schematics of electron-light-matter interactions in different consolidated and emerging forms of electron beam spectroscopies. In conventional CL/EELS (Fig. 1a), two-dimensional (2D) CL/EELS maps are acquired by raster-scanning the beam over the specimen. A 2D map is constructed for each emitted light wavelength (CL) or electron energy loss (EELS) in what has been termed hyperspectral imaging. A key aspect of the CL and EELS excitation mechanism is that the specimen is polarized by time-varying electric fields produced by the moving electron, similar to the effect of an optical pulse. The spatial extent of the radial electric field around the electron trajectory is shown in Fig. 1e in reduced units. The field decays evanescently at large distances, as described by the modified Bessel functions $K_0$ and $K_1$. As an example, Fig. 1f shows the time evolution of the radial and axial fields for a 30 keV electron at a position 5 nm away from the trajectory. The electron creates a single electromagnetic field cycle within a few hundred attoseconds. The corresponding frequency spectrum is shown in Fig. 1g and has energies with significant weight in the 0-30 eV spectral region, the precise range depending on the electron kinetic energy.[2,3] The electron thus acts as a broadband source of optical excitation (i.e., its electromagnetic field covers a wide spectral range), with a spatial resolution limited by the extent of its evanescent field (~0.5-10 nm, depending on electron energy and detection frequency). The EELS spectra are determined by the work done by each of the frequency components of the electron field acting on the polarized material, and probed through the ensuing losses experienced by the electron. Part of this work transforms into radiation emission (i.e., CL). An intuitive relation between EELS/CL and optical extinction/scattering can be rigorously established.[4] Importantly, CL and EELS involve incident (produced by the electron) and scattered (induced by interaction with the sample) electromagnetic fields with well-defined phase relations, as well as broadband, ultrafast, nanometer-precision attributes that are currently being unfolded through advances in these spectroscopies. Additionally, the cascade decay of excitations triggered by the primary electron can produce incoherent CL emission, for example by recombination of electron-hole pairs in semiconductor (nano)structures or the radiative decay of excited color centers in insulators.[5,6,7]

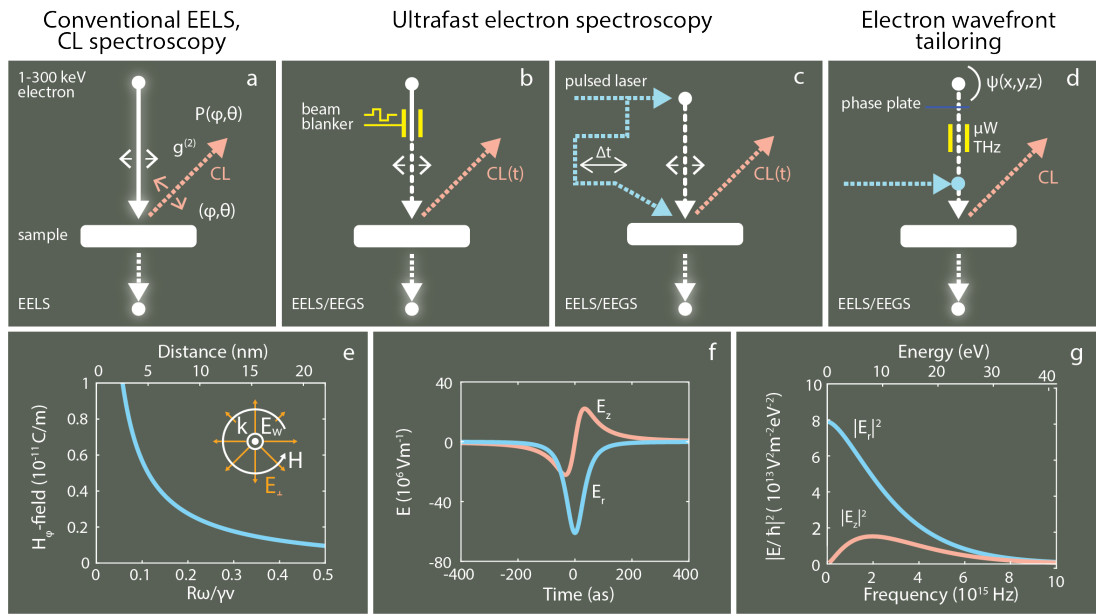

**Figure 1 | Electron-light-matter interactions**. **a** Conventional electron spectroscopy. The electron energy-loss spectrum (EELS) or the electron-induced (angle- and polarization-resolved) light emission (CL) is analyzed. **b,c** Ultrafast electron microscopy. Electron pulses are generated by an electrostatic beam blanker (b) or by photoemission from a pulsed-laser-driven cathode (c). Pulsed electron beams enable time-resolved imaging. When the specimen is optically excited, dynamic energy loss/gain (EELS/EEGS) and CL spectra can be acquired, enabling pump-probe spectroscopy to be performed by tuning the delay between laser and electron pulses on the sample (c). **d** Tailoring the electron wave function. Phase plates, microwave/THz cavities and intense optical near fields control the electron wave function $\psi(x,y,z,t)$. The shaped electron wave functions create unique ways to control coherent electron-matter excitations in advanced CL and EELS/EEGS. **e**. Spatial extent of the radial electric field around the electron trajectory in reduced units ($R$: radius, $\omega$: angular frequency, $\gamma$: Lorentz contraction factor, $v$: electron velocity). The top horizontal scale shows the distance $R$ for 30 keV electrons at a free-space wavelength of 800 nm. **f** Time evolution of the radial and axial electric fields for a 30 keV electron at a position 5 nm away from the trajectory. **g** Corresponding frequency/energy spectrum of the field intensity.

In the past few years, coherent CL and EELS spectroscopies have undergone revolutionary advances. Angle- and polarization-resolved CL have been introduced to fully characterize the state of the emitted CL light (Fig. 1a). Ultrashort electron pulses have been created through pulsed-laser photoemission in the electron cathode, laying the foundations of ultrafast electron microscopy (UEM) and time-resolved CL microscopy (Fig. 1b,c). As a consequence of this development, synchronized light and electron beam excitation of the specimen permits pump-probe CL or EELS spectroscopy to be performed and electron energy-gain processes can now be observed under pulsed sample excitation. Furthermore, several new approaches have been developed that shape the single-electron wave function in the spatial domain using elements placed along the electron beam path, or in the time domain using the interaction of the electron beam with suitably shaped optical fields (Fig. 1d).

In Section 2 of this review we briefly summarize the state-of-the art in the theory of CL, EELS and light-assisted UEM. We focus on coherent interactions related to the primary electron beam and do not discuss the incoherent processes that results from relaxation processes in the secondary electron cascade In Section 3 we describe highlights in CL/EELS nano-optics research enabled by technical developments in recent years. We focus on work that has led to new fundamental insights in coherent optical excitation of plasmons, plasmon polaritons, dielectric resonances and phonons in nanomaterials, as well as in the excitation of single-photon emitters. We also provide a comparison of the technical requirements and specifications of EELS and CL for nanophotonics research. In Section 4 we describe recent developments in ultrafast EELS and CL microscopy. In Section 5 we highlight recent advances in the shaping of electron wave functions in the spatial and time domains for time/space-structured electron-matter interactions. We conclude in Section 6 by providing an outlook on novel fundamental studies that are enabled by the new developments in coherent EELS/CL spectroscopy. We intend this review to stimulate further developments in these highly exciting novel areas of research, further exploiting the potential of CL, EELS and UEM for the investigation of nanoscale optical phenomena.

## 2. Theoretical description of coherent electron interactions with nanoscale optical excitations

*EELS and CL spectroscopies*

Fundamentally, both EELS and CL are excitation spectroscopies, commonly applied to reveal the strength of individual optical modes, as excited by the field displayed by the electron. Importantly, this field is coherent in the sense that it generates an induced field that is phase-locked to it, and for example the CL light emission in metals or photonic structures maintains phase coherence with the electron-generated field. In fact, the EELS probability is intimately related to the local density of photonic states (LDOS),[8] which is defined as the combined electric-field intensity of all normalized photonic modes as a function of light frequency and position in space, and in a more practical way, it is proportional to the decay rate of an optical emitter placed at that position.[9] In fact, it has been shown that one can reconstruct the LDOS from tomographic EELS measurements.[10] In a similar way, the CL intensity is related the radiative component of the LDOS (i.e., the component that is proportional to the radiative emission rate from the noted emitter). Interestingly, the resonance line shape can differ between CL and EELS spectra, essentially as a result of different frequency-dependent weighting functions accompanying the Lorentzian resonance profile, depending on whether one probes it in the near-field (through EELS or light absorption) or far-field (through CL or elastic light scattering).[4,11]

Intuitive insight into the coherent interaction of electron beams with photonic structures is gained by describing each moving electron as a classical point charge, whose electromagnetic field is interacting with the sample, as a result of which an induced field is produced. The latter encompasses far-field components, which represent scattering into emitted light (CL), and evanescent waves that act back on the electron producing energy loss (EELS). In this classical description, the electron supplies an external current density, and because we are interested in performing spectroscopy, it is useful to Fourier-transform it in time and decompose it in frequency components as $\mathbf{j}(\mathbf{r}, \omega)$ for each sample position $\mathbf{r}$.

By solving Maxwell's equation in the presence of structured materials, which enter through their frequency-dependent and spatially-dependent complex dielectric functions, the electron current, treated as a classical external source, permits us to obtain the resulting induced electromagnetic field $\mathbf{E}^{\mathrm{ind}}(\mathbf{r}, \omega)$. For CL spectroscopy, the induced far-field directly yields the emitted energy spectrum (through the Poynting vector), or equivalently, upon dividing by the photon energy $\hbar\omega$, the CL photon emission probability $\Gamma^{\mathrm{CL}}(\omega)$. In other words, the far electromagnetic field produced upon interaction of a moving charge (the electron) with the sample, which is obtained as the solution of the classical Maxwell equations, is time-Fourier transformed and its spectral decomposition interpreted as the probability of emission of photons as a function of their energy. While this interpretation emphasizes the quantum nature of the emitted photons, which is not really necessary to understand CL, it becomes essential to understand EELS. Indeed, for EELS, we can obtain the probability $\Gamma^{\mathrm{EELS}}(\omega)$ from the work done by the electron current, $-\mathrm{Re}\{\int d\mathbf{r}\, \mathbf{j}(\mathbf{r}, -\omega) \cdot \mathbf{E}^{\mathrm{ind}}(\mathbf{r}, \omega)\}$, divided again by $\hbar\omega$, and treating each frequency component separately. We note the quantum nature of the obtained spectra, emphasized by the introduction of Planck's constant $\hbar$: both CL and EELS are quantum-mechanical processes, in which individual photons or energy loss events are detected with a well-defined final electron energy, rather than an average classical far field or slowing down of the electron. Nonetheless, the prescription of treating each frequency component of the electron current separately, and ultimately dividing the emitted or absorbed energy by $\hbar\omega$, produces results in agreement with full quantum mechanical descriptions.[2] It is remarkable that this type of semi-classical analysis permits writing the EELS and CL probabilities in terms of the macroscopic electromagnetic response of the sample, and this in turn as a simple, yet generally accurate approximation as a function of the local material permittivities; one can also readily incorporate nonlocal dispersion effects in terms of momentum-dependent response functions along directions of translational symmetry (e.g., in the bulk of a material), while inclusion of these effects in arbitrary geometries lacking translational symmetry requires a more involved description of the constitutive relations, such as provided by first-principles simulations.

Following the above classical prescription, closed-form expressions of the EELS and CL probabilities have been derived for several simple geometries using analytical methods.[12,13,14] In particular, it is instructive to consider a simple polarizable point particle of polarizability $\alpha(\omega)$, for which, using CGS units, the EELS and CL probabilities per unit frequency reduce to:[2]

$$\left. \begin{array}{c} \Gamma^{\text{EELS}}(\omega) \\ \Gamma^{\text{CL}}(\omega) \end{array} \right\} = \frac{4e^2\omega^2}{\pi\hbar^2 v^4 \gamma^2} \left[ \frac{1}{\gamma^2} K_0^2\left(\frac{\omega R}{v\gamma}\right) + K_1^2\left(\frac{\omega R}{v\gamma}\right) \right] \times \left\{ \begin{array}{c} \text{Im}\{\alpha(\omega)\} \\ (2\omega^3/3c^3)|\alpha(\omega)|^2 \end{array} \right. \quad (1)$$

where the Lorentz factor $\gamma = 1/\sqrt{1-v^2/c^2}$ accounts for relativistic effects at the electron velocity $v$ ($= 0.33c$ and $0.70c$ at typical SEM and TEM beam energies of 30 keV and 200 keV, respectively). Note that the dependence on beam-particle distance $R$ is described by the modified Bessel functions $K_m$, and that $R$ is in fact normalized to the characteristic distance $v/\omega\gamma$, which determines the extension of the evanescent field associated with the passing electron, and in turn, the spatial resolution when imaging the excitation modes of the particle.[1,15] More precisely, the above probabilities decay as $K_m^2 \propto e^{-2\omega R/v}/\sqrt{R}$ at large separations and diverge at small distances as $K_0 \propto \log R$ and $K_1 \propto 1/R$. This scaling with $R$ therefore limits spatial resolution by $v/\omega$ in the regime of exponential decay, while only recoil, the physical width of the focused electron beam, and quantum-mechanical interactions impose a limit to spatial resolution at short distances; an accuracy as small as 3 nm is experimentally found for CL.[16] Additionally, these expressions reveal the same dependence of EELS and CL on the particle polarizability as the optical extinction and scattering cross sections, respectively, thus supporting the intuitive concept that EELS accounts for all loss channels of electron-sample energy transfer, whereas CL corresponds only to losses that result in the emission of radiation. As expected, the optical theorem[17] ($\text{Im}\{-1/\alpha(\omega)\} \geq 2\omega^3/3c^3|\alpha(\omega)|^2$) directly implies $\Gamma^{\text{EELS}} \geq \Gamma^{\text{CL}}$. The relation between EELS/CL and extinction/scattering can be similarly extended to arbitrarily shaped nanoparticles in the quasi-static regime.[4]

In general, samples in actual experiments require a numerical solution of Maxwell's equations, for which various approaches have been developed, including the boundary-element method (BEM)[18] and the useful MNPBEM implementation,[19] the discontinuous Galerkin time-domain method,[20] the discrete dipole approximation,[21,22] the finite-difference in the time-domain method (FDTD),[23,24] generalized Mie theory,[25,26] and innovative approaches coupling these equations to the quantum electron-wave-function dynamics.[27] In a complementary theoretical effort, advanced spectral-image processing techniques have been recently applied to obtain tomographic reconstructions of the spatial extent of localized optical modes[10,28] (see Section 3).

The above analytical and numerical methods rely on a dielectric description of the sample to obtain EELS and CL probabilities, an approach that has general applicability and can be used in combination with local, frequency-dependent dielectric functions to cope with plasmons, phonon-polaritons, and excitons in most samples. In particular, surface phonon polaritons, which have been recently the subject of intense research because of their potential as mid-infrared modes with long lifetimes,[29] can now be probed by EELS thanks to recent advances in TEM instrumentation (see below), and the local dielectric approach generally leads to sufficiently accurate calculations of the obtained loss spectra.[30] A quantum-mechanical treatment of the fast electron further permits simulating lateral momentum transfers (i.e., inelastic electron distributions as a function of their deflection angle) and the reshaping of the lateral electron wave function upon interaction with excitations in the specimen.[2,31] It should be noted that when the sample is structured at length scales comparable to the Fermi wavelength of the involved materials, ranging from less than 1 nm in noble metals to 10s of nm in highly doped graphene, spatial dispersion and quantum finite-size effects become important and provide substantial corrections beyond local response, thus demanding the use of more sophisticated first-principle-based methods, that lie beyond the scope of the present review.

*Interaction with optical fields and ultrafast microscopy*

The development of UEM has demanded a quantum treatment of the electron to explain the acquired EELS spectra.[32,33] In fact, the interaction between short electron and laser pulses interacting through the mediation of the sample provides a handle to manipulate the electron wave function along the beam direction. The use of laterally extended beams adds intriguing effects resulting from the interplay between lateral and parallel wave function components, as recently shown through the demonstration of orbital angular momentum transfer between light and electrons.[34,35]

Although energy-momentum mismatch severely limits the free-space interaction between electrons and light, electrons can efficiently couple to evanescent optical fields that are produced by optical excitation of a material structure, as these can produce additional momentum transfer that break the noted mismatch. A semiclassical treatment of these interactions has been successfully used to explain recent experiments,[32,36] in which the quantum-mechanic electron wave function evolves in the presence of a classical external light field $2\text{Re}\{\mathbf{E}^{\text{ext}}(\mathbf{r})\,e^{-i\omega t}\}$. We note that this semiclassical description where the electron dynamics is described quantum mechanically and the light field classically accounts for electron-light interaction when the light is indeed classical (e.g., supplied through the coherent states of a pump laser), but it cannot be extended to describe the self-interaction of the electron through its own induced field (i.e., for EELS and CL), which requires a quantum description of the electron-sample interaction.[2] For monochromatic light of frequency $\omega$, the incident electron wave function $\psi^{\text{inc}}(\mathbf{r},t)$ picks up inelastic components $\psi_\ell(\mathbf{r},t) \approx \psi^{\text{inc}}(\mathbf{r},t) J_\ell(2|\beta|) e^{i\ell \arg\{-\beta\} + i\ell\omega(z-vt)}$, where $J_\ell$ is the Bessel function of the first kind and corresponding to the absorption ($\ell > 0$) or emission ($\ell < 0$) of $\ell$ photons by the electron, where the coupling integral[37]

$$\beta(x,y) = \frac{e}{\hbar\omega} \int dz\, E_z^{\text{ext}}(\mathbf{r}) e^{-i\omega z/v}, \qquad (2)$$

presented here for an electron moving along the $z$ direction, captures the interaction with the optical field as a function of lateral position $(x,y)$. A simple extension of this description has been formulated for pulsed electrons and light,[32,38] with the interaction still governed essentially by $\beta(x,y)$. In its simplicity, Eq. (2) captures the essence of the interaction between swift electron beams and light in laser-illuminated samples, and consequently, it is widely used in the explanation of many experiments.[34,39] We stress that the coupling only involves the electric field component along the electron trajectory, and conversely, the electron can mainly sample this component of the optical field. This integral directly explains the vanishing interaction between electrons and free-space light, as $\omega/v$ exceeds any light wave-vector component arising from $E_z^{\text{ext}}$, therefore rendering $\beta = 0$ in the absence of a material structure. It is reassuring to note that the CL probability of Eq. (1) can be directly obtained from Eq. (2) when applying it to a point particle and setting the incident light intensity to the LDOS times the field per vacuum photon mode.[37]

## 3. Coherent electron-matter interactions

*Exciting localized plasmons*

Noble metal nanoparticles possess strong localized plasmon resonances that render them as ideal building blocks in nanometer-scale photonic architectures, triggering different research areas within the field of nanophotonics during the last decade. The relatively strong coupling between energetic electrons and plasmons has been exploited in numerous EELS and CL studies, taking advantage of the unparalleled spatial resolution of these techniques. While the first EELS experiments on nanometer-sized metal (and semiconductor) nanoparticles were carried out several decades ago,[40,41,42,43] it took until 2007 for improvements in energy and spatial resolution of EELS to enable direct visualization of plasmonic modes in Ag and Au nanoparticles (Fig. 2a).[44,45] A very high spatial imaging resolution of ~λ/40 was achieved, far below the optical diffraction limit. Following these pioneering papers, EELS has been extensively used to identify and map plasmonic modes in a wide variety of resonant plasmonic nanostructures at high spectral resolution (see Ref. [46] and references therein). Importantly, EELS probes electron losses associated with both radiative and non-radiative processes, so in contrast to far-field optics, it can be used to map not only electric dipole modes, but also quadrupoles and higher-order modes that do not, or only weakly, couple to far-field radiation.

CL spectra and images of plasmonic modes on Ag nanoparticles were first reported using 200 keV electrons in a TEM.[47] Several years later it was found that 30 keV electrons in a SEM also create efficient CL signals on plasmonic structures, and spatial maps of plasmons in Au nanowires were presented.[48,49]. Over the past decade, a number of CL studies have identified plasmonic modes in a wide variety of resonant nanostructures, nearly all carried out using SEM-CL systems (Fig. 2b). Plasmon resonances studied by CL range from the ultraviolet (e.g., in Ga and Al) to the visible and near-infrared (e.g., in Ag, Cu and Au) spectral domains (see Refs. [6], and [7] and references therein), while recent advances in EELS spectrometers and monochromators also enable the study of mid-infrared modes.[50] Therefore, EELS and CL are key techniques to tackle novel emerging research areas in plasmonics, including transdimensional materials (between 2D and 3D),[51] aluminum plasmonics,[52,30,53,54] doped metal oxides[55] or refractory transition metal nitrides.[56]

*Angle- and polarization-resolved CL*

In CL, the angle dependence of the emitted light intensity can be measured over a wide angular range. Measurements are performed by projecting the emitted light that is collected by a parabolic mirror onto a CCD camera using a suitably designed optical path. Such angular measurements explore an important degree of freedom to probe details of localized plasmon resonances.[57,58,59] For example, in CL experiments on a 150-nm-diameter Au disk, the excitation of multiple radiative resonant modes leads to interference in the far field, resulting in distinctly shaped angular emission profiles.[60] Differences in the angular distribution of the CL emission can also yield information on plasmon mode symmetry.[61] This interference is a direct demonstration of the coherent nature of electron-beam excitation of multiple resonant plasmonic modes by a single electron. Furthermore, in CL polarimetry the four Stokes parameters can be derived from six independent polarization-filtered measurements, enabling spatial and angular mapping of the full polarization state of CL.[62] Exploiting this effect, the localized electron excitation of specially designed plasmonic geometries can be used to create tailored angular and polarization states of light in the far field.[63] Vice versa, angle-resolved imaging and polarimetry enable partitioning of the CL spectra in incoherent and coherent components.[63,64]

*Plasmon CL/EELS tomography*

Three-dimensional information on the sampled specimen can be acquired from tomographic reconstructions using multiple CL or EELS measurements performed under different electron incidence angles. Using this concept, the three-dimensional (3D) plasmonic modes were determined using EELS in Ag nanocubes and nanoparticles by taking a series of measurements under different tilt angles[65,66] (Fig. 2c). A similar technique was used to retrieve the LDOS in plasmonic Ag dimers.[10,28] Complementarily, 3D CL tomography was shown to retrieve the 3D distribution of the resonant modes of metallo-dielectric nanoparticles through the analysis of a large number of CL measurements on identical particles at different angles (Fig. 2d).[67]

*Exciting surface plasmon polaritons*

In parallel to the broad research area of noble metal nanoparticles that possess localized plasmon resonances, the study of surface-plasmon polaritons (SPPs) that propagate at the interface between a metal and a dielectric has gained great interest. SPPs are highly confined waves with wavelengths that can be much smaller than free space waves at the same frequency. High-energy electrons serve as ideal point sources for SPPs on a planar metal-dielectric interface.[68,69] They create femtosecond plasmon wavepackets that propagate at the interface with a gradually decaying intensity due to Ohmic dissipation in the metal.[3] Pioneering angle-resolved EELS experiments revealed the dispersion of SPPs on thin Al films.[70] Angle-resolved EELS has been used in general for systems in which translational invariance along one or several directions makes momentum rather than spatial resolution relevant. This has been applied to bulk structures, for example for band gap determination,[71] 2D systems such as plasmonic surfaces[72] and 1D systems such as carbon nanotubes.[73] In resonant nanostructures, CL measurements of the plasmonic standing waves provide a unique way to probe the SPP wave vector at a given frequency, so that from a range of measurements over a broad frequency band the dispersion relation can be determined.[48,74,75,76] Complementarily, angle-resolved CL measurements probe the SPP wave vector in periodic plasmonic crystal structures, from which the plasmonic band structure can be derived (Fig. 2e).[77] EELS is also used to study optical properties of anisotropic materials. It particular, it was employed to study plasmons in graphite[78] or hexagonal boron nitride[79] and their curved forms, such as onion fullerenes and nanotubes,[80] including those formed with transition metals dichalcogenides.[81] With successive advances in energy resolution, finer lowenergy details, such as band gaps[82] or excitonic lines could be measured in nanotubes.[83] The anisotropy of these systems further allow to detect different forms of optical excitations such as the Dyakonov mode.[84]

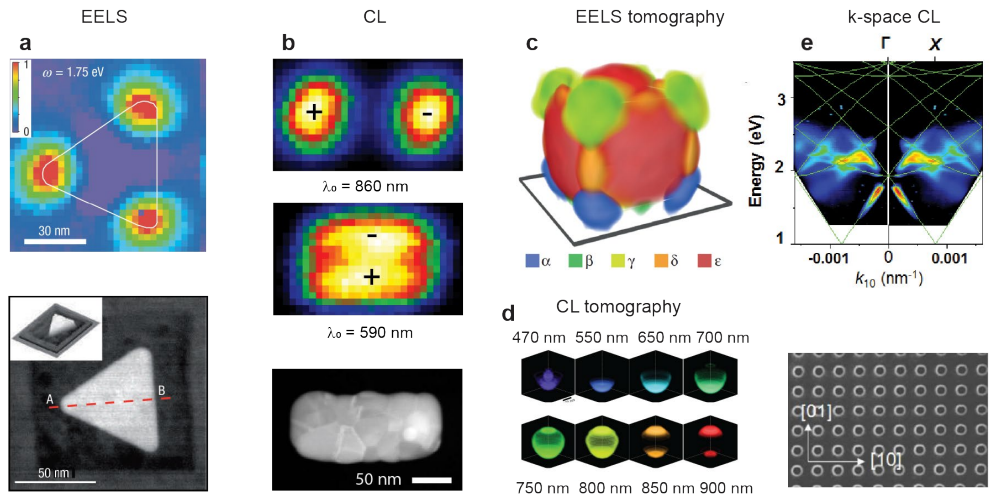

**Figure 2 | EELS and CL spectroscopy of plasmonic nanostructures. a** (top) EELS map at corner plasmon resonance energy (1.75 eV) for a triangular Ag particle probed with 100 keV electrons. (bottom) TEM image of the same structure. From Ref. [45]. **b**. (top) CL images of transverse and longitudinal resonant plasmonic modes at 590 nm and 860 nm of a Au nanowire, acquired with 30 keV electrons. (bottom) TEM image of the same structure. From Ref. [49]. **c** 3D tomographic reconstruction using EELS (300 keV) of the field distributions of five different resonant plasmonic modes indicated by different colors (energy 2.2-3.5 eV) in a 100-nm Ag nanocube. From Ref. [65]. **d** 3D tomographic reconstruction using CL (30 keV) of the plasmonic field distributions for different light wavelengths in a 230-nm-diameter polystyrene/Au-nanoshell geometry. After Ref. [67]. **e** (top) Dispersion diagram derived from angle-resolved CL data on a square array of holes (pitch 600 nm) drilled in a Ag film. The CL intensity is plotted along the Γ-X direction for p-polarized emission. Green lines indicate band-folded calculated SPP dispersion curves without the holes. (bottom) SEM image of the hole array. From Ref. [77].

*Exciting resonant dielectric nanostructures*

Electron beam excitation provides a unique way to probe resonant modes of dielectric (non-plasmonic) nanoparticles. Here, the fast electrons directly couple to the localized polarizable molecular bonds inside the dielectric, setting up displacement polarization currents governed by resonant electric and magnetic Mie modes. Using this principle, EELS was used in an aloof geometry (i.e., with the beam passing just outside the material) to identify Mie modes in silica spheres in the far-ultraviolet spectral range.[85] Further EELS measurements on dielectrics are scarce, partly due to limitations imposed by broadening of the zero-loss peak in thick specimens. In contrast, CL is ideally suited to identify the Mie modes of silicon nanodisks in the visible-to-near-infrared spectral range, and their spatial modal distributions were imaged at a resolution far below the optical diffraction limit.[86] CL has enabled imaging of the cavity modes in dielectric photonic crystals, for which the angular CL emission profiles reflect their photonic band structure.[87] Electron beams also strongly couple to travelling optical waves: infrared CL measurements have probed the modal field distributions of TE and TM polarized modes in Si photonic crystal waveguides,[88] and most recently, topological Si photonic crystals.[89] In a related context, EELS was used to detect photonic Bloch modes in porous $Al_2O_3$ membranes[90] and Si photonic structures.[91]

*Imaging phonons*

A key advance in EELS is the development of microscope systems with an energy resolution down to <10 meV, bringing within reach an entirely new field of vibrational electron spectroscopy.[92] This is the result of the development of bright cold field-emission electron sources, new efficient electron monochromator designs, spectrometers, improved aberration-corrected beam optics and improved resolution in the electron energy detectors. Taking advantage of this high energy resolution it has recently been demonstrated that surface and bulk phonons in MgO cubes can be mapped at the nanoscale using their characteristic EELS peaks (Fig. 3).[93] Interestingly, the low-energy vibrational states can be excited using aloof excitation, as the relative contribution of low frequencies in the electron excitation increases for larger distances away from the beam path.[94] Among them, excitations such as surface phonons are of prime interest in IR nanophotonics, playing a similar role as surface plasmons in nanophotonics at higher frequencies. Surface phonons, in addition, may present competetive advantages compared to plasmons because, for a proper choice of materials, they can exhibit quality factors much larger than those predicted for plasmons.[95]

The interaction of fast electrons with phonons under aloof excitation can be described in similar terms as the excitation of surface plasmons,[30] where EELS closely map the photonic density of states. The description for penetrating trajectories, where bulk phonons are excited becomes more involved.[93,96]. Nevertheless, upon reciprocal-space removal of the surface phonon signals, atomically-resolved maps of bulk phonons can be resolved.[97] Interestingly, the vibrational EELS spectra measured on MgO showed both energy loss and gain components, reflecting both the creation and annihilation of phonons by the electron beam, with the EELS gain/loss peaks described by a Boltzmann distribution, confirming the inelastic nature of the electron scattering by phonons in thermal equilibrium.[98,99]

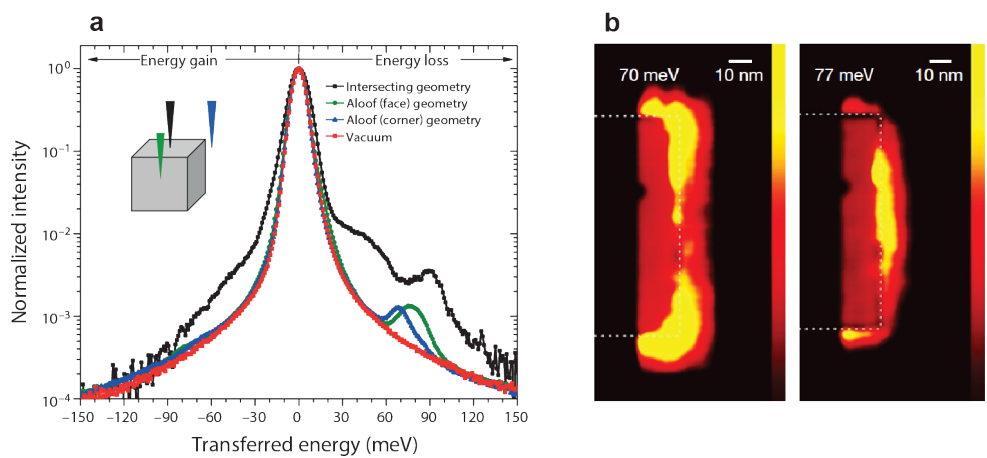

**Figure 3 | EELS phonon microscopy. a** EELS spectra taken on a single 100-nm MgO cube. **b** EELS maps showing the spatial distribution of corner (left, 70 meV) and face (right, 77 meV) vibrational modes on the edge of the MgO cube. From Ref. [93].

*CL photon statistics*

The unique way in which electrons excite optical materials creates special fingerprints in the photon statistics of the CL light that is emitted. When single-photon emitters, such as for example nitrogen-vacancy (N-V) centers in diamond, are excited by energetic electrons, the emission itself has no phase relation with the incoming electron, as intermediate energy relaxation processes precede the excitation of emitter. However, TEM-CL measurements of the $g^{(2)}$ second-order autocorrelation function for CL photon emission from individual N-V centers in diamond[100] or point defects in hexagonal BN[101] show strong anti-bunching, corresponding to the excitation of a single photon emitter (Fig. 4a). Complementarily, $g^{(2)}$ measurements on multiple centers in TEM-CL showed strong photon bunching, in strong contrast to the photoluminescence observations. This is due to the fact that a single electron can excite multiple energetic secondary electrons that can in turn create many optical excitations, leading to the emission of a bunch of photons within a time determined by the excited-state lifetime of the emitter.[102] This renders the measurement of $g^{(2)}$ as a powerful tool for the determination of lifetimes at high spatial resolution.[103] Photon correlation measurements were also carried out in SEM-CL (Fig. 4b), in which pulsed excitation using electrostatic blanking creates an additional degree of freedom to control the statistics of the excitation process. These measurements enabled the direct determination of the efficiency of materials excitation by the primary electron[104] and have also enabled unravelling the relative excitation and emission probabilities of semiconductor nanostructures by analyzing $g^{(2)}$ data with a statistical model.[105]

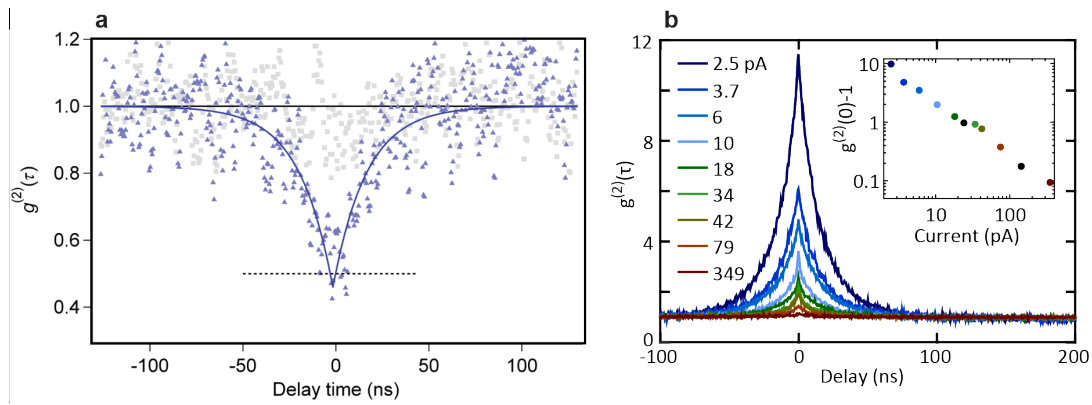

**Figure 4 | Photon bunching and anti-bunching in incoherent CL**. Spectrum of the second-order autocorrelation function $g^{(2)}$ of CL emission. **a** TEM-CL on diamond nanoparticles with a single embedded N-V center (80 keV electrons, 570-720 nm spectral band) showing photon emission anti-bunching. From Ref. [100]. **b** SEM-CL on InGaN quantum wells for different beam currents showing bunching. From Ref. [104].

**Table I | Characteristics of state-of-the-art EELS and CL spectroscopy.**

|  | EELS | CL |
|---|---|---|
| Instrument configuration | (S)TEM with electron energy detector | S(T)EM with CL light collection and analysis system |
| Optical processes probed | Sum of non-radiative and radiative processes | Radiative processes |
| Energy range | > 0.03 eV ($\lambda$<30 $\mu$m) | $\lambda$=200-900 nm using Si CCD detector<br>$\lambda$=900-1700 nm using InGaAs detector |
| Spectral resolution | ~10 meV (2 nm bandwidth at $\lambda$=500 nm) | ~0.1 nm (0.3 meV at 2 eV) |
| Electron probe size | TEM-EELS: < 0.1 nm | TEM-CL: < 0.1 nm<br>SEM-CL: ~1-10 nm |
| Specimen thickness | <50 ~ 100 nm | No limit |
| Normalization of spectral intensity | Normalizing by the full spectral integral. | Using transition radiation as a reference, for example from a planar Al surface. |

## 4. Ultrafast EELS and CL

*Ultrafast electron microscopy*

The low-energy gain effects observed in vibrational EELS spectroscopy (Fig. 3) result from the annihilation of thermal phonons. Much larger energy gains can be observed in EELS when the specimen is excited by spatially and temporally overlapping laser and electron pulses. This idea was proposed in Ref. [106], theoretically elaborated,[37] and eventually demonstrated in experiments on carbon nanotubes and silver nanowires under visible light excitation (Fig. 5a).[36] In these photon-induced near-field electron microscopy (PINEM) experiments, a ladder of energy loss and gain peaks was observed.[36] The data represent the strong nonlinear coupling between the electron and photon fields, in which many energy quanta are exchanged for each electron in the beam. Key to the efficient electron-light interaction described here is the use of optical near fields, as plane waves and electrons cannot exchange energy because of lack of energy-momentum conservation, as was already mentioned in Section 2. In contrast, optical near-fields contain evanescent components that can fulfill the required conservation laws, thus making PINEM possible. Interestingly, by mapping the energy gain spectra across a specimen that is resonant with the incident light, the optical near-field of the resonance can be reconstructed, as recently shown by probing the optically excited SPP Fabry-Perot modes on a Ag nanowire,[107] and a subsequent demonstration of 20 meV resolution based on varying the light frequency around those modes.[53]

A key element in the energy gain experiments is the synchronous excitation of a specimen by laser and electron pulses.[108] In these experiments a beam from a femtosecond pulsed laser is split in two parts: one to excite the specimen, and one to induce photoemission from the electron cathode, which leads to the generation of electron bunches. Typical electron pulse durations are around 1 ps and the average number of electrons per pulse can range between less than 1 and over 1000, depending on laser pulse fluence and cathode settings such as temperature and extraction voltage.

Pulsed-laser-driven cathodes were first used in electron diffraction to investigate, for example, transient phase transformations,[109] as well as structural relaxations.[110] Their use in time-resolved electron microscopy has rapidly grown in recent years. Recent detailed studies of the PINEM effect have shown that the quantized energy sidebands are populated in a quantum-coherent way according to the Rabi oscillations between the light and electron states.[39] These experiments confirmed theoretical predictions,[32] and the quantized energy gain/loss transitions are so strong that they can lead to a near-complete depletion of the initial electron energy state (Fig. 5b). Importantly, the quantum superposition of the excited electron ladder states results in coherent shaping of the electron wavepacket in momentum space, creating a train of attosecond electron pulses in the time domain. Tailoring of the electron wavefunction will be further discussed in Section 5.

*Ultrafast CL*

The first use of a pulsed photoemission microscope in CL spectroscopy was reported in Ref. [111]. Later, using 200 fs laser pulses to generate 10-ps electron pulses (10 keV), the carrier dynamics in GaAs nanostructures was probed with a spatial resolution of 50 nm (Fig. 8).[112] The ultrafast pulsed geometry also enables pump-probe spectroscopy with laser and electron beams as pump and probe, or vice versa, opening up an entirely new research area of ultrafast excited matter spectroscopy using CL and EELS.

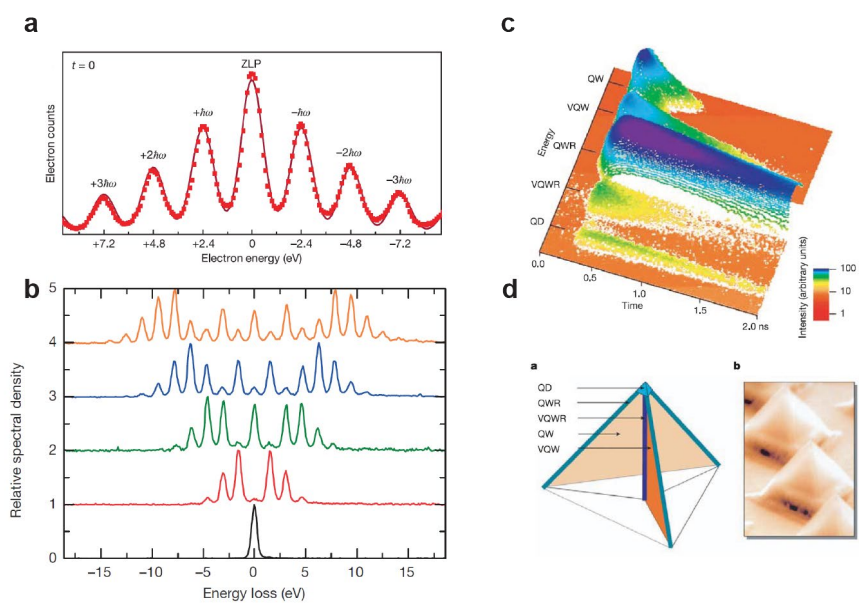

**Figure 5 | Ultrafast electron microscopy a** Energy gain/loss spectroscopy and quantum coherent manipulation of electron energy distributions. **a** Energy spectrum of 200 keV electrons after interaction with carbon nanotubes that are simultaneously excited with a pulsed laser (200 fs, $\hbar\omega$= 2.4 eV), showing three energy gain and three energy loss quanta. From Ref. [36]. **b** Energy spectrum of 120 keV electrons after interaction with an optical near field around a Au needle (3.5 ps, $\hbar\omega$= 1.6 eV). Data are shown for different pulse energies increasing from bottom to top. Energy gain/loss peaks are quantized by the photon energy and distributed according to Rabi oscillations coupled among the ladder states, creating an attosecond train of electron pulses. From Ref. [39]. **c** Streak

images of CL emission from quantum dot, quantum wire and quantum well structures in a InGaAs/AlGaAs micropyramid (1.52-1.90 eV, 90 K) taken using a 10 keV pulsed electron beam generated by photoemission with a 200-fs laser pulse ($\lambda$=266 nm). **d** Schematic of sample geometry and SEM image. Tops of pyramids are separated by 5 $\mu$m. From Ref. [112].

## 5. Tailoring the electron wave function

As is clear from Sections 3 and 4, there have been many advances in EELS, EEGS and CL spectroscopy in the past years that have led to new insights in multiple aspects of electron-matter-light interactions. These experiments have relied on state-of-the-art (pulsed) electron microscopes employing "conventional" electron beams that have phase fronts similar to plane waves, with the spatial and temporal coherence determined by the source and electron column geometries and settings. Recent exciting new developments concern tailoring the electron wave function itself, both in the spatial and time domains. A first example was already discussed above, where PINEM experiments on Au nanowires created electron pulses composed of a train of attosecond pulses.[39]

*Time domain*

It is well known that electrons can elastically scatter from light fields by the ponderomotive force, such as for example in the Kapitza-Dirac effect,[113,114] in which two counterpropagating optical waves configure a light grating that can diffract a passing electron wave. However, these elastic interactions are too weak for practical applications in beam shaping. The evanescent optical components produced when light interacts with material boundaries provide an efficient way to enhance such interactions. For example, in the inelastic Smith-Purcell effect, electrons propagating above a grating interact with their induced electromagnetic surface waves, resulting in the generation of light in the visible spectral range and a concomitant loss in electron energy.[115,116] A similar diffraction effect is achieved when electrons interact with periodic field patterns of plasmonic standing waves excited by optical pumping.[117] Complementarily, in the "inverse Smith-Purcell effect" electromagnetic surface waves generated by optical pulses can accelerate electrons, as most recently demonstrated for near-infrared pulsed laser excitation.[118,119] Inelastic electron-near-field-light scattering effects provide a unique way to tailor the wave function of the (single) electron itself, enabling coherent control over electron-light-matter interaction.

Additional approaches have been recently explored to tailor the electron wave function with far-field radiation pulses. For example, by placing a pulsed microwave source inside the electron column, single-electron pulses could be compressed and tailored.[120] The oscillating microwave cavity field pulse is synchronized with the electron pulse such that it acts differently on the front and rear parts of the electron pulse, effectively widening the energy spread and shrinking the spatial distribution (pulse width) of the electron pulse in the direction along the beam (Fig. 6a,b). In this way, the electron pulse can be made shorter than the original laser pulse generating it. Using this concept, a train of electron pulses compressed by THz pulses was used to follow the excitation and relaxation of THz-excited Ag butterfly antennas in the time domain at a time resolution of ~5 fs, by using an ingenious scheme in which the time-varying electric near field of the antenna controlled the streaking of the electron pulse.[121,122]

*Spatial domain*

Initial experiments in shaping the electron beam in the spatial domain used either rotated superimposed graphene sheets[123] or holographic gratings[124] to induce a singular spiraling phase to the electron wave, generating vortex electron beams that carry orbital angular momentum. Since then, numerous methods for producing various phase-shaped beams have been proposed (see Ref. [125] for a review). For instance, using advanced phase plates, electrons carrying quantized amounts of orbital angular momentum up to $\pm100\hbar$ per electron have been created (Fig. 6c).[126] An alternative technique to create electron vortex beams uses a magnetic needle placed in the electron beam path.[127] Recently, direct transfer of angular momentum from an optical beam to the electrons has been demonstrated.[34]

The research area of structured beams is now emerging further, and some initial applications have been demonstrated. Vortex beams can probe circular magnetic dichroism, similar to X-ray magnetic circular dichroism spectroscopy,[124] but now with the prospect of atomic resolution. Similarly, it has been predicted that the chirality of plasmonic systems could be determined using a vortex beam.[128,129] Aside from vortex beams, several other beam geometries have been realized using structured surfaces.[130,131,132] The first application of a phase-shaped beam in nanophotonics was in the measurement of the symmetry of the electric field amplitude of localized modes of a plasmonic nanorod, which is not possible in a direct way using EELS or CL, as these techniques probe the field intensity rather than the amplitude.[133] The same concept can be expanded to other symmetries, in order to reveal the symmetries of a broad range of coherent excitations. The use of ultrathin materials, such as for example graphene, which interact strongly with both electrons and light, can provide further control over electron-light-energy exchange processes.[134] More generally, phase information about the coupling to the nanophotonic system is encoded in the electron wave front after interaction with the sample; it can be unfolded by acquiring and analyzing the reciprocal (angular) space images. However, this information is generally lost in conventional EELS setups, in which collection up to high detection angles averages out phase variations, as early envisioned by Kohl[135] and Ritchie and Howie.[136] Ptychographic techniques (i.e., those reconstructing the phase by using information hidden in the diffraction pattern acquired for every point of a scan) constitute a generalization of the concepts of structured beams and permit retrieving phase information without requiring dedicated beam phase shapes adapted for each problem, as experimentally demonstrated in recent experiments.[137]

In a recent exciting development, a programmable electron phase plate was demonstrated that is composed of an array of cylinders in which the electric field is individually manipulated (Fig. 6d).[138] This enables tunable control over the beam geometry that is reconstructed by interference after the electron is transmitted through the array.[138] This concept, which can be expanded further to large array sizes, holds great potential as a unique way to control beam shape in a detailed way.

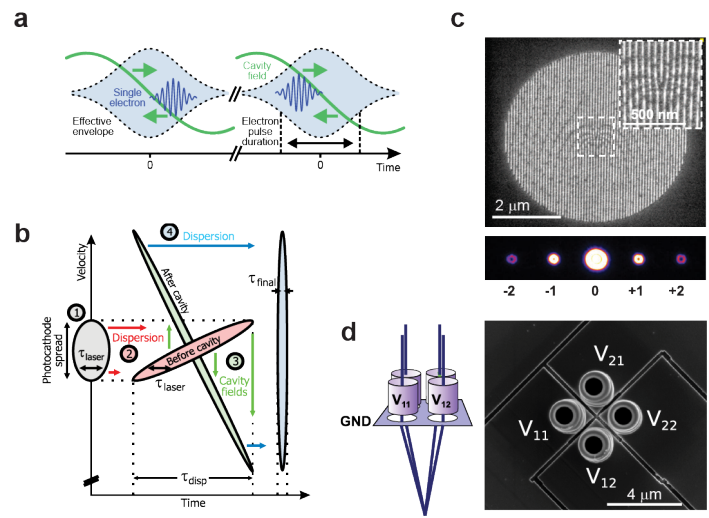

**Figure 6 | Tailoring the electron wave function in the temporal and spatial domain**. **a** The time-varying electric field in a microwave pulse modifies the electron wavepacket to a desired shape. **b** Schematic of electron pulse in the velocity and time domain, before and after passing through a microwave cavity. From Ref. [120]. **c** Phase plate composed of a transmission grating with fork dislocations. Transmitted electron waves carry discrete orbital angular momenta with topological charges indicated at the bottom. From Ref. [126]. **d** Programmable tunable phase plate composed of a 2×2 array of cylindrical electrodes that individually control the phase of 4 electron beams that recombine in the far field to form a programmable interference pattern. From Ref. [138].

The creation of electron wave functions with tailored spatial, temporal and angular momentum distributions is strongly connected to developments in structured illumination and pulse shaping that have proven very powerful in (super-resolution) optical microscopy and spectroscopy. Similar applications in electron microscopy are opening up at a fast pace. In fact, one could imagine that temporally and spatially structured (multi-color) illumination sources interacting with electrons could provide even further control over the optical shaping of electron wavepackets.

## 6. Future directions

As is clear from the many examples reviewed above, the EELS/CL community is very lively with new discoveries continuously being made. Based on the described developments we envision several notable trends.

*Time resolution*

Developments in ultrafast laser-driven cathodes are continuing and electron pulses as short as 200 fs were recently demonstrated,[139] bringing time-resolved studies of hot electron and electron-phonon relaxation in solids within reach. So far, the generation of single pulses short enough to directly probe plasmon relaxation (~10 fs) has remained elusive. Trains of attosecond pulses have been realized, as described above. An exception is a recent point-projection technique, in which plasmon dynamics of the optically excited electron cathode itself was imaged with a spatial resolution of 20 nm and a time resolution of 25 fs, thus getting closer to the real-time characterization of plasmon dynamics.[140] An alternative proposal is to retrieve the sub-cycle dynamics of plasmons by using interference between the field of an specially designed plasmonic metamaterial lens excited by a fast electron and a plasmonic field of interest.[141,142] Recent developments with electron diffraction (without spatial resolution) have demonstrated sub-fs resolution in pulse trains, paving the way to sub-optical cycle temporal resolution in EELS/CL.[143] As a benchmark in this context, photoemission electron microscopy (PEEM) has been shown to render <10 nm spatial resolution through electron imaging, accompanied by ~10 fs resolution associated with pump/probe delay in two-photon photoemission (see below).[144,145]

We note that temporal information on the electron excitation processes can also be derived from EELS data using Fourier analysis methods.[146] Ultrafast transient behavior may also be resolved by studying the effect of attosecond forces exerted on plasmonic nanoparticles induced by swift electrons.[147,148]

We note that recent advances have been triggered by the development of high-brightness Schottky guns. We anticipate that the search for even brighter guns will play an important role in the years to come, while significant progress has been already made with the conception of the first time-resolved cold-field emission gun.[149]

Finally, we note that recent work shows how electrostatic beam blankers placed in the electron column can now deliver electron pulses as short as 30-90 ps[150,151] thus enabling CL lifetime imaging of a broad range of materials at very high spatial resolution. The advantage of these blankers is that they can be easily integrated in the electron column. Another exciting new development is the use of an electrostatic beam blanker driven by a laser-driven photoconductive switch, which may produce electron pulses as short as 100 fs.[152]

*Energy resolution*

A key new development in EELS instrumentation is the demonstration of energy resolution below 10 meV. This enables a wide range of high-resolution phonon/electron spectroscopy studies in bulk materials, as well as at surface and interfaces.[92] An interesting parallel development is the use of microwave fields created by RF cavities that are integrated in the electron column to tailor the electron pulse and energy resolution. A new design shows that 200-fs electron pulses may be created with an energy width below 500 meV.[153] Using a sequence of GHz RF cavities even promises energy resolutions down to 20 meV, far better than the energy spread in the source itself. These designs come within a factor 10 of the Heisenberg limit of the variance in beam energy and position. In principle, such new designs could simplify the EELS instrumentation, as they require a much simpler microscope column, although the required (high) beam currents have not yet been realized. The use of pulsed electron beams in EELS opens up the use of time-of-flight energy loss analysis.

We note that electron energy-gain spectroscopy (EEGS) has been proposed[37] as a way to combine the excellent energy resolution in the frequency of the external laser with the atomic resolution provided by electron beams. The promised EELS mapping of optical excitations with sub-meV resolution still remains as an experimental challenge. A recent development has used this principle to map narrowly separated plasmon standing waves in the spectra of long silver wires with 20 meV resolution.[53]

*Structured electron beams as quantum electron probes*

As discussed in Sections 4 and 5, it has now become possible to tailor the electron wave function in both the spatial and temporal domains. Spatially structured beams provide a new degree of freedom in spatially-resolved excitation and will enable advanced studies on symmetries in plasmonic excitations. Temporally structured beams can provide coherent control over optical excitations and effectively perform pump-probe spectroscopy with both the pump and probe encoded in the same pulse, thus effectively granting access to sub-optical-cycle dynamics.

An exciting aspect of these developments is that they pursue to control the electron wave function itself. From a fundamental perspective, electron beams are used as quantum probes, with the electron microscope operating as a quantum instrument using well-prepared initial states that can be entangled with materials excitations. In particular, when a CL photon or a loss signal is recorded in EELS, one makes sure that one quantum of excitation has been produced on the sample. This enables exciting studies on correlations in time and space, as well as real-time studies of excited states by their spectral, spatial and diffraction signatures. A key element in these studies is to obtain optimum sensitivity, extracting as much information as possible from the smallest possible number of electrons, aiming to achieve unity detection quantum efficiency. A fundamental question arises, whether quantum information can be encoded in the electron spectrum in the form of coherent superpositions of optically excited states. This also raises the question how electrons can be used to probe quantum aspects in a specimen, how the collapse of the electron wave function to an observable eigenstate can be controlled, and whether this could be exploited in (scalable) quantum technology. Quantum measurements may also enable new forms of microscopy in which the electron-sample interaction is probed with one part of the electron wave function, while a weak entangled part is interacting with the specimen, keeping electron-induced degradation to a minimum.[154]

*Incoherent electron-matter interactions*

Although this review focuses mostly on coherent light-matter interactions, it is important to note that several new developments are taking place that involve incoherent processes and that provide unique new insights into quantum optical phenomena and condensed-matter energy landscapes. Recently, the very high spatial resolution of electron spectroscopy has been employed to investigate transition-metal dichalcogenides (TMDCs) with unique (semi-)metallic and semiconducting properties. For example, using EELS the spin-orbit energy splitting in $MoS_2$ and $MoSe_2$ was measured[155] and low-energy hyperbolic phonon polariton modes were observed in hexagonal BN.[156] Additionally, CL measurements have revealed sulfur impurity inhomogeneities in $MoSe_2$ flakes.[157] By embedding a TMDC layer in a heterostructure geometry, the effective capture range of electron-induced excitations can be strongly enhanced, and the specimen stability improved, enabling the observation via CL of strain-induced lateral bandgap variations in $WSe_2$ monolayers at length scales below the optical diffraction limit.[158]

*Complementary techniques*

Aside from the focus on EELS and CL in this review, two other techniques involve (ultrafast) electron-matter-light interactions. In particular, PEEM, which renders images constructed from photoelectrons emitted from a specimen, can render a spatial resolution down to <10 nm, as determined by the electron optics. Field enhancement caused by, for example, plasmonic hotspots locally enhance the photoemission intensity, so that PEEM images directly probe plasmon fields.[159] The spectral and temporal resolution is determined by the exciting photon beam properties, allowing few-femtosecond-resolution observation of plasmon interference dynamics when the exciting light is administered in a pump-probe configuration.[160] Also, the polarization and phase of the incoming light beam can be adapted for studying the plasmon symmetry. As a striking example, PEEM has been recently used to investigate the ultrafast dynamics of plasmonic vortices.[145] More details on PEEM and its relation with EELS, EEGS and CL can be found in Ref. [161]. We note that the related ultrafast point-projection electron microscopy (see above) also presents a way to directly image the plasmon intensity distributions at high spatial and temporal resolution.[140]

In scanning tunneling microscopy (STM), inelastic electron tunneling produces luminescence with spectral features that render information on the sampled materials and structures. Optical spectra are collected with the atomic spatial resolution offered by the STM, although the details of the tip morphology, which are usually unknown, add a certain degree of uncertainty on the imaged optical modes. Nevertheless, this technique has proven to be useful for the investigation of electro-optical molecules[162,163] and plasmonic fields.[164]

*Compact microscopes*

The new fundamental developments also raise the question whether further improvements in electron microscope designs can be made by critically reconsidering features that have historically been developed. Obviously, the use of lower electron energies simplifies the design of the columns. Introducing novel beam shaping concepts as described above can reduce the need for correctors, further simplifying the electron column. Altogether, the distinction between TEM and SEM geometries may vanish for some applications. Ultimately, compact ("table-top") microscope designs may become possible taking advantage of the new electron beam shaping concepts.

## 7. Conclusions

In summary, the research field of electron beam spectroscopy for nanophotonics has grown into an exciting research field, enabled by many recent technical advances in CL and EELS spectroscopies during recent years. Electron beams offer materials excitations in the optical spectral range at atomic (EELS) and nanometer (CL) spatial resolution, and provide unique access into the optical response of nanophotonic structures at very high time and energy resolution. The newest high-resolution EELS systems have opened up a new field of electron-excited phonon microscopy. Aside from these coherent excitation processes, incoherent excitation of semiconductors and single quantum emitters creates photon bunching and anti-bunching, providing further detailed insights into electron-matter interactions.

The high degree of spatial and temporal control that can now be achieved over electron beams anticipates the development of exciting new applications in quantum coherent control. Electron pulses are quantum probes that can be potentially entangled with materials excitations, raising fundamental questions if they can carry quantum information to create entirely new forms of electron microscopy with entangled beams. Overall, the many new insights in electron beam spectroscopy for nanophotonics promises many exciting new discoveries in electron-light-matter interaction in the coming years.

**Acknowledgements**

We gratefully acknowledge the assistance of Sophie Meuret and Toon Coenen in preparing this review. Future visions described in this paper partly originate from presentations and a panel discussion session at the workshop Electron Beam Spectroscopy for Nanophotonics (EBSN) held in Sitges, Spain, during October 25-27, 2017. We thank the workshop participants for providing their insights; in particular the discussion panelists Joanne Etheridge, Ido Kaminer, Claus Ropers, and Johan Verbeeck. The Dutch part of this work is part of the research program of the Netherlands Organization for Scientific Research (NWO); the French part has received support from the French State through the National Agency for Research under the program of future Investment EQUIPEX, and TEMPOS-CHROMATEM with the reference ANR-10-EQPX-50; the Spanish part is supported by MINECO (MAT2017-88492-R and SEV2015-0522), the Catalan CERCA Program, and Fundació Privada Cellex. This project has received funding from the European Research Council (ERC) under the European Union's Horizon 2020 research and innovation programme (grant agreements No. 695343 and 789104). The authors declare the following competing financial interest: A.P. is co-founder and co-owner of Delmic BV, a company that produces commercial cathodoluminescence systems.

**ORCID**
Albert Polman 0000-0002-0685-3886
Javier García de Abajo 0000-0002-4970-4565
Mathieu Kociak 0000-0001-8858-0449